\documentclass[aps,superscriptaddress,twocolumn,showpacs,preprintnumbers,amsmath,amssymb]{revtex4}
\usepackage{pst-all} 
\usepackage{multido}
\usepackage[dvips]{graphicx}
\usepackage{eucal}
\usepackage{bm} 
\usepackage{float}
\usepackage{color}

\newcommand{\Red}[1]{#1}

\newcommand{\npart}{n_\mathrm{part}}

\newcommand{\comment}[1]{}

\begin{document}
\title{
Effects of fluctuations on the initial eccentricity
from the Color Glass Condensate
in heavy ion collisions
}

\medskip

\author{Hans-Joachim Drescher}
\affiliation{
Frankfurt Institute for Advanced Studies (FIAS),
Johann Wolfgang Goethe-Universit\"at,
Max-von-Laue-Str.~1, 60438  Frankfurt am Main, Germany
}
\author{Yasushi Nara}
\affiliation{
Institut f\"ur Theoretische Physik,
Johann Wolfgang Goethe-Universit\"at,
Max-von-Laue-Str. 1, 60438  Frankfurt am Main, Germany
}

\begin{abstract}
We introduce a modified form of the Kharzeev-Levin-Nardi (KLN)
approach for nuclear collisions. The new {\em ansatz} for the unintegrated
gluon distribution function preserves factorization, and the
saturation scale is bound from below by that for a single nucleon. It also
reproduces the correct scaling with the number of collisions at high
transverse momentum. The corresponding Monte Carlo implementation
allows us to account for fluctuations of the hard sources (nucleons)
in the transverse plane. We compute various definitions of the
eccentricity  within the new approach, which are relevant for
the interpretation of the elliptic flow. Our approach predicts
breaking of the scaling of the eccentricity with the Glauber
eccentricity at the level of about 30\%.
\end{abstract}


\maketitle
\section{Introduction}

The elliptic flow $v_2$ is one of the most important observables in
high energy non-central nucleus-nucleus collisions at
RHIC~\cite{RHICexperiments}. It is very sensitive to both
the initial condition in the overlap zone, and to the subsequent evolution
of the hot matter created in these collisions~\cite{Ollitrault:1992bk}.
Predictions from ideal hydrodynamics with an initial condition from a
Glauber type model and with a short thermalization time $\tau \leq 1$
fm/c are compatible with the elliptic flow data at low transverse
momentum at RHIC~\cite{hydroreview} energies.  This is one of the
strongest indications of early thermalization and nearly perfect-fluid
evolution of the quark-gluon plasma at RHIC.
 
It was found in Ref.~\cite{Hirano:2005xf} that ideal hydrodynamics
with the initial condition taken from the Color Glass Condensate (CGC)
approach~\cite{HiranoNara04} overestimates the measured elliptic flow
at RHIC. Since hadronic dissipative effects were taken into account by
a hadronic transport model~\cite{jam}, additional dissipation in the
quark-gluon phase would be required.
The reason for the strong flow is that the KLN approach~\cite{KLN01}
predicts a larger eccentricity of the initial gluon distribution in
the transverse plane than the Glauber
model~\cite{Hirano:2005xf,Drescher:2006pi}.
A recent calculation within the classical Yang-Mills (CYM)
approach on the lattice~\cite{LV06}
gave somewhat smaller values than predicted by the KLN
model. Nevertheless, the eccentricity of the ``overlap'' region in
coordinate space obtained from CYM still exceeds that predicted by the Glauber
``wounded nucleon'' model for soft processes.

The eccentricity $\varepsilon$ is defined in the transverse
plane $\bm{r}_\perp=(r_x,r_y)$ perpendicular to the beam axis by
\begin{equation}
\label{eq:ecc}
\varepsilon =
\frac{\langle r_y^{\,2}{-}r_x^{\,2}\rangle}
     {\langle r_y^{\,2}{+}r_x^{\,2}\rangle}~.
\end{equation}
The average may be taken with respect to either energy or particle
density (or entropy, if rapid thermalization is assumed).  In
the ideal hydrodynamical limit~\cite{Bhalerao05}, elliptic flow only
depends on the density and on $\varepsilon$. Therefore, it is very
important to know the initial conditions to extract properties of hot
and dense matter created in nucleus-nucleus collisions.

This paper is a continuation of Ref.~\cite{Drescher:2006pi}, where we
computed the eccentricity in the initial stage of heavy ion collisions
within the KLN approach, and showed that the large eccentricity
predicted by the $k_\perp$-factorization approach is generic, as it
does not depend on the details of the unintegrated gluon distribution
function (uGDF) such as the presence of leading-twist shadowing and an
extended geometric scaling window~\cite{DHJ,KKT03,Braun:2000bh}.

The predictions of the CGC framework are consistent with RHIC data for
the centrality, energy and rapidity dependence of the multiplicity of
charged hadrons if the square of the saturation scale (averaged over
the transverse plane) is defined to be proportional to the number of
participants \cite{KLN01,Drescher:2006pi,Baier:2002bt,Armesto:2004ud}.
In this paper, we shall consider in more detail the relation between
the different definitions of the saturation scales (see also the short
note in Ref.~\cite{Drescher:2006pi}).

The eccentricity could of course fluctuate for many reasons, and it
was recently argued that event-by-event flow fluctuations can affect
the elliptic flow obtained via correlation methods
~\cite{Miller:2003kd,Manly:2005zy,Voloshin:2006gz,Bhalerao:2006tp}.
In order to explore such fluctuation effects in the CGC picture, we
employ a Monte-Carlo implementation of the KLN approach which accounts
for fluctuations of the high-rapidity ``hard sources'' (nucleons) in
the transverse plane.  This leads naturally to a factorized version of
the KLN approach, where the uGDFs depend only on the properties of a
single nucleus.  We analyze the effects of the fluctuations on the
eccentricity by using different definitions of the eccentricities
~\cite{Miller:2003kd,Manly:2005zy,Voloshin:2006gz,Bhalerao:2006tp}.
This is relevant to the study of the effects of fluctuations on $v_2$
~\cite{Aguiar:2001ac,Hama,Andrade:2006yh,Zhu:2005qa}.

\section{Gluon production in the KLN approach}

In the $k_\perp$-factorization approach~\cite{GLR83},
the number distribution of produced gluons is given by
\begin{eqnarray}
  \frac{dN_g}{d^2 r_{\perp}dy}&=&
  \frac{4N_c}{N_c^2-1} \int^{p_\perp^\mathrm{max}}\frac{d^2p_\perp}{p^2_\perp}
  \int^{p_\perp} \frac{d^2 k_\perp}{4} \;\alpha_s  \\
  & \times&   \phi_A(x_1, (\bm{p}_\perp+\bm{k}_\perp)^2/4)\;
              \phi_B(x_2, (\bm{p}_\perp{-}\bm{k}_\perp)^2/4)
  \nonumber
 \label{eq:ktfac}
\end{eqnarray}
with $N_c=3$ the number of colors.  Here, $p_\perp$ and $y$ denote the
transverse momentum and the rapidity of the produced gluons,
respectively. The light-cone momentum fractions of the colliding gluon
ladders are then given by $x_{1,2} = p_\perp\exp(\pm y)/\sqrt{s}$,
where $\sqrt{s}$ denotes the center of mass energy and $y$ is the
rapidity of the produced gluon.  We set
$p_\perp^\mathrm{max}$ such that the minimal saturation scale
\Red{$Q_{s,\min}(x_{1,2})$}
in the above integration is
$\Lambda_{QCD}=0.2$~GeV (see, however, below).
The KLN approach~\cite{KLN01} employs the following uGDF:
\begin{equation}
\label{eq:uninteg}
  \phi(x,k_\perp^2;\bm{r}_\perp)\sim
  \frac{1}{\alpha_s(Q^2_s)}\frac{Q_s^2}
   {{\rm max}(Q_s^2,k_\perp^2)}~,
\end{equation}
where $Q_s$ denotes the saturation momentum at the given momentum fraction
$x$ and transverse position $\bm{r}_\perp$. The overall normalization
is determined by the multiplicity at midrapidity for the most central
collisions. 

In Refs.~\cite{KLN01,HiranoNara04,Hirano:2005xf,Drescher:2006pi},
the saturation scale for nucleus $A$ is taken to be proportional
to the density of participants, $n^A_\mathrm{part}(\bm{r}_\perp)$,
which in turn
is obtained from the thickness functions $T_A$ and $T_B$:
\begin{eqnarray}
 n^A_\mathrm{part}(\bm{r}_\perp) &= &
    T_A(\bm{r}_\perp + \bm{b}/2) \nonumber\\
     &\times& (1-(1-\sigma_{NN}T_B(\bm{r}_\perp - \bm{b}/2)/B)^B)~.
      \label{eq:npart}
\end{eqnarray}
$\bm{b}=(b,0)$ is the impact parameter vector in the transverse plane
and $\sigma_{NN}$ is the nucleon-nucleon inelastic cross section.

The density of participants is not a universal quantity which depends
only on the properties of a single nucleus.  In the following we
propose a modified form of the KLN approach where factorization is
manifest and show that the $Q_s^2\sim\npart$ definition is in fact a good
approximation for computing the multiplicity of gluons. Other
quantities such as the energy density, however, do behave somewhat
differently.

\section{Factorized KLN formula for nucleus-nucleus collisions}

\begin{figure}[tb]
\begin{center}
\includegraphics[width=8cm,clip]{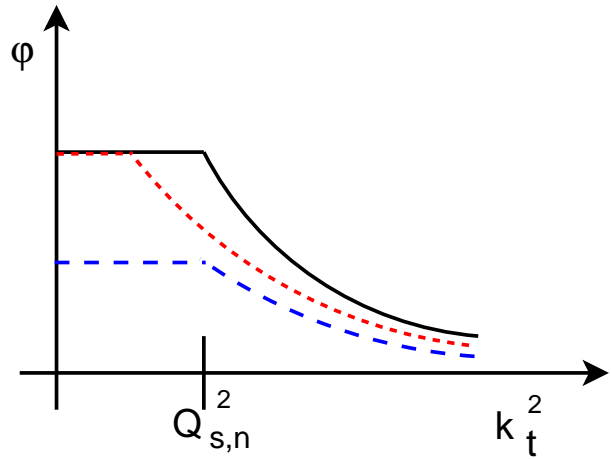}
\caption{(Color online)
Schematic sketch of different constructions of the uGDF.
The full line corresponds to the uGDF of a single nucleon,
the other lines to that of a nucleus at its surface.
The dotted line corresponds to the the definition from
Eq.(\ref{eq:wrong}) while the dashed line is for 
Eq.~(\ref{eq:right}).
}
\label{fig:phi}
\end{center}
\end{figure} 

Generically, it is reasonable to define the squared saturation scale
to be proportional to the density of nucleons. However, a problem
arises near the edge of a nucleus where the density is small. There,
the thickness function $T_A$ is small: only for {\em some}
configurations of nucleons do we actually find a nucleon at that
position. Let us denote this probability of finding at least one
nucleon at a given transverse coordinate as $p_A$. For configurations
where indeed there is one single nucleon at the given ${\bf r}_\perp$,
the uGDF should be that of a
nucleon: $\phi=\phi(Q_{s,n}^2)$.  However, the average $\phi_A({\bf
r}_\perp)$ would then be
\begin{equation}
\phi_A = p_A \phi(Q^2_{s,n})
\label{eq:right}
\end{equation}
rather than
\begin{equation}
\phi_A = \phi(Q^2_{s,n}p_A)~.
\label{eq:wrong}
\end{equation}
This is illustrated in Fig.~\ref{fig:phi}. The essential point is that one
has to average the uGDF itself, and not its argument $T_A$.
Expressing the uGDF
in terms of $T_A$, the relevant variable would be $T_{1A}$, which is the
density of nucleus $A$ averaged only over those configurations with at
least one nucleon
at a given transverse position. This quantity can be obtained
by averaging over the thickness functions of $i\geq 1$ nucleons,  
\begin{equation}
T_{1A} \equiv  \langle t_A \rangle_{i\geq 1} =
  \frac{\sum_{i=1} p_{A}(i) ~ t_A(i) }{\sum_{i=1} p_A(i)}
         = \frac{T_A}{p_A},
\end{equation}
where $t_A(i)$ is the thickness function for $i$ nucleons in a row.
This allows us to construct an uGDF in the following way,
\begin{equation}
\phi_A = p_A ~ \phi \left( \frac{T_A}{p_A} \right)~. \label{eq:phi_factorized}
\end{equation}
Here, and in the following, we use a short-hand notation for the
dependence of $\phi$ on some density $\rho$: $\phi(\rho)\equiv
\phi(Q_s^2(\rho))$. The uGDF (\ref{eq:phi_factorized}) respects
factorization, since it depends only on the properties of a single nucleus.
The saturation scale is then parameterized as
\begin{equation}
 Q^2_{s,A}(x,\bm{r}_\perp) = 
   2\,{\rm GeV}^2\left(\frac{T_A(\bm{r}_\perp)/p_A(\bm{r}_\perp)}{1.53}\right)
                       \left(\frac{0.01}{x}\right)^\lambda,
		        \label{eq:qs}
\end{equation}
$p_{A}$ can be taken from the Glauber definition of $\npart$:
\begin{equation}
p_A = (1-(1-S T_A /A)^A) \label{eq:pA}.
\end{equation}
In this case, $S$ is a sampling area and not the
nucleon-nucleon inelastic cross section as in the Glauber
model. Nevertheless, these quantities are related to each other, 
since $S$ should be on the order of the area of the nucleon.
One could fix its value in such a way that one 
recovers the saturation scale of a nucleon in the limit $T_A
\rightarrow 0$, since
\begin{equation}
\lim_{T_A \rightarrow 0} \frac{T_A}{p_A} = \frac{1}{S} . 
\end{equation}
For our simulations we will however use the value of the inelastic
cross section at RHIC energy, since then the results are more directly
comparable to the original KLN ansatz. The saturation scale of a
nucleon is then equal to $Q_{s,n}^2=2.0~\mathrm{GeV}^2/(4.2\times
1.53)=0.31~\mathrm{GeV}^2$.  We refer to this approach as the
factorized KLN (fKLN) model.

Next, we explore the relation to the original KLN ansatz, where 
the saturation momentum is defined via $\npart$.
The transverse density of produced gluons can roughly be
expressed analytically in terms of the saturation scales as~\cite{KLN01}
\begin{equation}
\frac{dN_g}{dyd^2\bm{r}_\perp} \sim   Q^2_{s_,\min}
     \left(2+\log\left(\frac{Q^2_{s,\max}}{Q^2_{s,\min}}\right)\right),
\end{equation}
\Red{
where, $Q_{s,\max}$ and $Q_{s,\min}$ denote the larger and
the smaller value of the two saturation scales in opposite nuclei at any
fixed position in the transverse plane.
}
It is easy to see that this function is homogeneous of order one in
both $Q_{s,A}^2$ and $Q_{s,B}^2$, or in the
corresponding densities which appear in the definition of the
saturation scales.
The $k_\perp$-factorization formula with the new definition of the
uGDF (fKLN approach) now
takes the form:
\begin{equation}
 \frac{dN_g}{dyd^2\bm{r}_\perp} \sim  \int \frac{d^2 p_\perp}{p_\perp^2} \int d^2 k_\perp 
p_A p_B \phi(T_A/p_A) \phi(T_B/p_B) . \label{eq:kt2}
\end{equation}
Due to the homogeneity, we can move the prefactors in
eq.~(\ref{eq:kt2}) into the arguments of the uGDF. If $S$ is the
nucleon-nucleon inelastic cross section $\sigma_{NN}$, we have
$n_{\mathrm{part},A} = T_A p_B$.  This leads to the original KLN definition of
the uGDF, where the saturation scales depend on the
participant densities:
\begin{eqnarray}
 \frac{dN_g}{dyd^2\bm{r}_\perp} &\sim&  \int \frac{d^2 p_\perp}{p_\perp^2} \int d^2 k_\perp 
 \phi(p_A p_B T_A/p_A) \phi(p_A p_B T_B/p_B) \nonumber \\
 &=&   \int \frac{d^2 p_\perp}{p_\perp^2} \int d^2 k_\perp 
 \phi(n_{\mathrm{part},A}) \phi(n_{\mathrm{part},B}) ~.
\end{eqnarray}

\begin{figure}[tb]
\begin{center}
\includegraphics[width=8cm,clip]{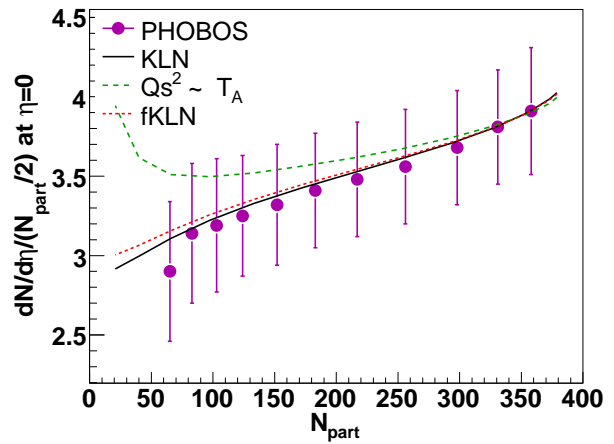}
\caption{(Color online)
Centrality dependence of the multiplicity for the different
assumptions of the uGDF.
Eq.~(\ref{eq:wrong}) gives a significant overestimate of the
multiplicity in the peripheral collisions.  The original KLN ansatz,
where $Q^2_s \sim \npart$ is very similar to the fKLN result which
uses Eq.~(\ref{eq:kt2}).  Data from PHOBOS~\cite{phobos}.  }
\label{fig:npart_TA_pA}
\end{center}
\end{figure} 
\begin{figure}[tb]
\begin{center}
\includegraphics[width=8cm,clip]{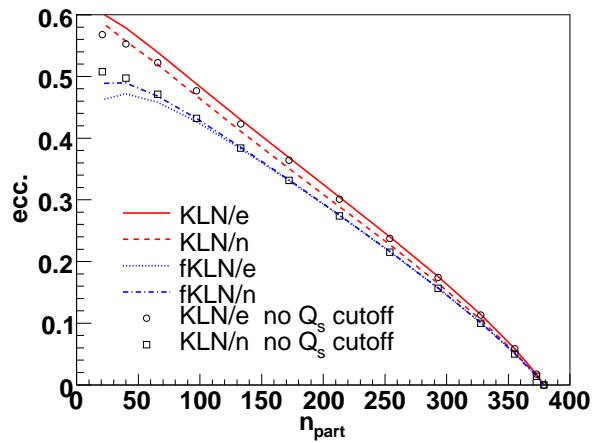}
\caption{\Red{(Color online) The initial eccentricity $\varepsilon$ at
    midrapidity as a function of the number of participants for
    200\,$A$\,GeV Au+Au collisions: KLN/e (KLN/n) corresponds to the KLN result using the
    energy (number) density as the weight in the definition of the
    eccentricity.  fKLN results follow the same notation.  The KLN
    results without a saturation scale cutoff are shown by symbols.  } }
\label{fig:ecc_pA}
\end{center}
\end{figure} 

The expression~(\ref{eq:kt2}) involving the new definition of the uGDF
can now be integrated numerically.  The
centrality dependence of the multiplicity at
midrapidity is shown in Fig. \ref{fig:npart_TA_pA}, assuming that the
hadron yield is proportional to the yield of produced gluons. For
comparison, we also show the result corresponding to the
definition $Q^2_s
\sim T_A$ in the $k_\perp$-factorization formula, which clearly
overshoots the data for peripheral collisions.
On the other hand, the fKLN approach is rather
similar to the original KLN approach.
This is due to the fact that the single-inclusive gluon cross section
is homogeneous of order one in both arguments, as discussed above.

In Fig.~\ref{fig:ecc_pA}, \Red{we compare the eccentricities obtained
from fKLN (eq.~(\ref{eq:ecc})) and KLN. We plot two cases, where we
use the transverse energy density $dE_\perp/dyd^2{\bf r}_\perp$ and
the number density as a weight in the definition of the
eccentricity.  } The eccentricity from fKLN is somewhat lower than
the one from KLN, especially for more peripheral collisions.  This is
due to the fact that the transverse energy distribution from the KLN
approach is homogeneous of order 3/2, because of another factor of
$Q_{s,\max}$~\cite{DMcL}:
\begin{equation}
\frac{dE_\perp}{d^2{\bf r}_\perp dy} \sim 
Q_{s,min}^2Q_{s,max}\;\log\frac{Q_{s,max}^2}{Q_{s,min}^2} .
\end{equation}
As a consequence, the energy density distributions in the transverse
plane from KLN and fKLN, respectively, are somewhat different.

If one uses the number density of produced gluons rather than their
energy density as a weight in the definition of the eccentricity, one
would expect that the fKLN form reproduces the KLN result, since the
same arguments hold as for the multiplicity. However, we have found
that the fKLN prediction for the eccentricity with the number density
as a weight is almost identical to the above result (which employed energy
density as a weight). This can be traced back to the above-mentioned
cut-off $Q_{s,min}=\Lambda_{\rm QCD}$. In the fKLN model no such
cutoff is needed as the minimal saturation momentum never drops below
the saturation momentum of a single nucleon, by construction. Rather,
the probabilities $p_A$, $p_B$ of encountering at least one nucleon
vanish at the surface of the overlap region.  Even though it does not
make much sense physically, technically we can set the minimal
saturation scale in the KLN model to zero. Then the eccentricity from
the KLN approach agrees well with the one from the factorized fKLN
form \Red{as seen in Fig.~\ref{fig:ecc_pA}.}

\section{Monte-Carlo implementation of the KLN approach}

According to the ideas of McLerran and Venugopalan, the gluon field of
a sufficiently large nucleus can be viewed as a classical Yang-Mills
field~\cite{MV}.  Particle production in high-energy hadronic
collisions can then be calculated by solving the Yang-Mills equations
in the forward light-cone with a boundary condition corresponding to
classical sources on its two branches. Real time
solutions of the classical Yang-Mills fields on the lattice
were obtained, for example, in Refs.~\cite{KV,Krasnitz:2002mn,Lappi}.
Those numerical solutions are usually averaged over many different initial
conditions (configurations of the color-charge sources).

Here, we calculate the thickness function used for determining the
saturation scale with a Monte Carlo method.  There are two advantages
of doing this.  First, the Monte-Carlo implementation naturally leads
to a definition of the saturation momentum in terms of the thickness
functions (which preserves factorization). Second, we can study the
effects of event-by-event fluctuations on the eccentricity within the
KLN approach. We only consider fluctuations of the positions of the
nucleons, which were previously studied in
Refs.~\cite{Miller:2003kd,Manly:2005zy,Voloshin:2006gz,Bhalerao:2006tp}
within the Glauber approach.
We note that in Ref.~\cite{Krasnitz:2002mn}, a similar method was
employed to impose a color neutrality condition for a each nucleon in
the classical McLerran-Venugopalan model in order to simulate
collisions of finite nuclei.

\subsection{Model}

Let us now construct a Monte Carlo model for gluon production for each
configuration of nucleons within the colliding nuclei (denoted as
MC-KLN). We first determine the positions of $A$ nucleons according to
the Woods-Saxon distribution. Correlations among the nucleons are
neglected at present but could in principle be taken into
account. They tend to reduce fluctuations somewhat~\cite{Baym:1995cz}.

The local density
of nucleons, $t_A$, is then obtained by counting nucleons within a
tube of radius $r_\mathrm{max} = \sqrt{S/\pi}$ around a given
transverse coordinate ${\bf r}_\perp$:
\begin{equation}
  t_A(\bm{r}_\perp) = \frac{\mathrm{number~of~nucleons}}{S}
\end{equation}
The area $S$ is the sampling area over which we count nucleons. The
average $\left< t_A(\bm{r}_\perp) \right>$ reproduces the convoluted
thickness function $\widetilde{T}_A(\bm{r}_\perp)$ with a sampling
profile $f({\bm r}_\perp)=\frac{1}{S}\Theta (r_{max}-|{\bm
r}_\perp|)$:
\begin{equation}
\left< t_A({\bm r}_\perp) \right> = \widetilde{T}_A({\bm r}_\perp) 
   = \int d^2{\bm r}'_\perp ~ T_A({\bm r}'_\perp)
                            ~ f(\bm{r}_\perp-{\bm r}'_\perp)
\end{equation}
In the limit $S\rightarrow0$, $\widetilde{T}_A$ equals $T_A$. This
limit can be taken in a Monte Carlo by the oversampling
(``test-particle'') technique, whereby the sampling area $S$ is
reduced by some factor $n$ and the number of nucleons is increased by
the same factor; each nucleon then carries a weight $1/n$.  (We found
that oversampling by a factor of $n=100$ gives stable results.)  It is
not appropriate to simply decrease $S$.  While this would reproduce
the mean value of $T_A$, it would also change the fluctuations in the
number of nucleons at a given position, since each nucleon would
contribute $n/S$ to the density. With the oversampling technique each
nucleon has a weight $1/n$ and therefore contributes $1/S$ to the
density, which is independent of $n$.  However, correlations between
neighboring points in the transverse plane are destroyed.  Therefore,
event-by-event observables cannot be obtained with this method; we
employed it only to test convergence of our MC implementation to the
``mean-field'' results presented above and for the computation of
averaged observables.  In the following simulations,
we choose $S=\sigma_{\rm inel}$ as in the mean-field fKLN approach above.

We then follow the same procedure to compute $t_B(\bm{r}_\perp)$,
thereby generating two configurations for nucleus $A$ and $B$,
respectively, at a specified impact parameter.  The saturation scale
at a given transverse coordinate is then given by
\begin{equation}
 Q^2_{s,A}(x,\bm{r}_\perp) = 
   2\,{\rm GeV}^2\left(\frac{t_A({\bm r}_\perp)}{1.53}\right)
                       \left(\frac{0.01}{x}\right)^\lambda
		        \label{eq:qs} ~,
\end{equation}
and similarly for nucleus $B$.
It is clear that this choice of the saturation scale only depends on
the properties of one nucleus, and therefore respects factorization.

For each generated configuration we apply
the $k_\perp$-factorization formula~(\ref{eq:ktfac})
to obtain the distribution of produced gluons in the transverse plane.
We then average over many collisions to compute
observables and their fluctuations.

Finally, we establish the relation between MC-KLN and fKLN.
The average uGDF which is effectively employed in the MC-KLN model
may be written as
\begin{equation}
\phi_A = \sum_{i=0}^A p_A(i) \phi(t_A(i))
       = \sum_{i=1}^A p_A(i) \phi(t_A(i))  ~,
\label{eq:phi_average}
\end{equation}
where we used $t_A(0)=0$.
The fKLN uGDF may be obtained from the approximation
where $t_A(i)$ is replaced by the average, $t_A(i) \to T_{1A}$:
\begin{equation}
\phi_A \approx \sum_{i=1}^A p_A(i) \phi(T_{1A})
       =  p_A \phi(T_A / p_A ),
\end{equation}
where $\sum_{i=1}p_A(i)=p_A$ was used.
%

\subsection{Results}

\begin{figure}[tb]
\begin{center}
\includegraphics[width=8cm,clip]{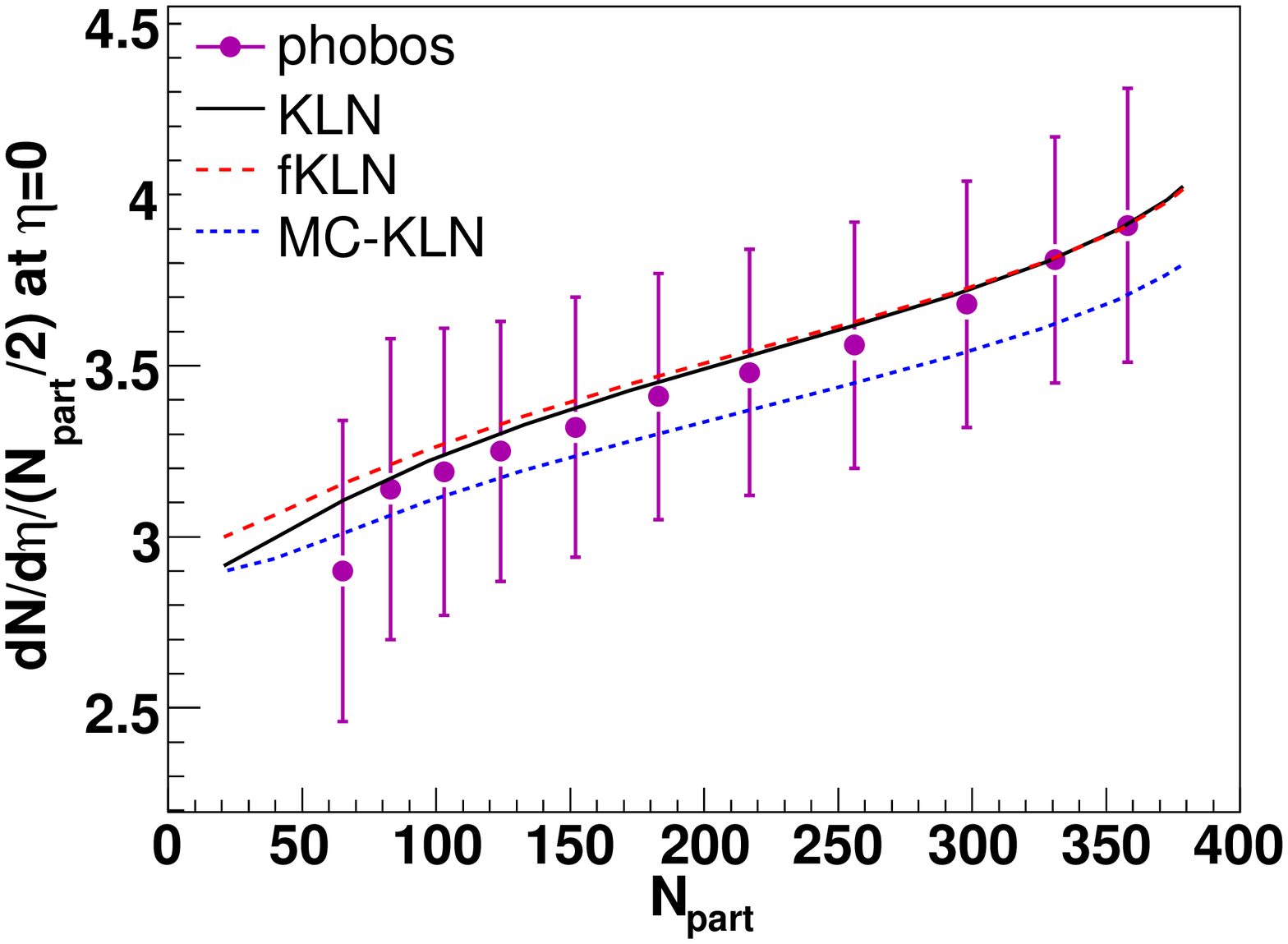}
\caption{(Color online)
Centrality dependence of the multiplicity in Au+Au collisions
for $\sqrt{s_{NN}}=200$~GeV for the fKLN and MC-KLN models.  All curves
have the same normalization, we adjusted the result for KLN to the two
most central bins. The data is from PHOBOS~\cite{phobos}.  }
\label{fig:npart_TA_MC}
\includegraphics[width=8cm,clip]{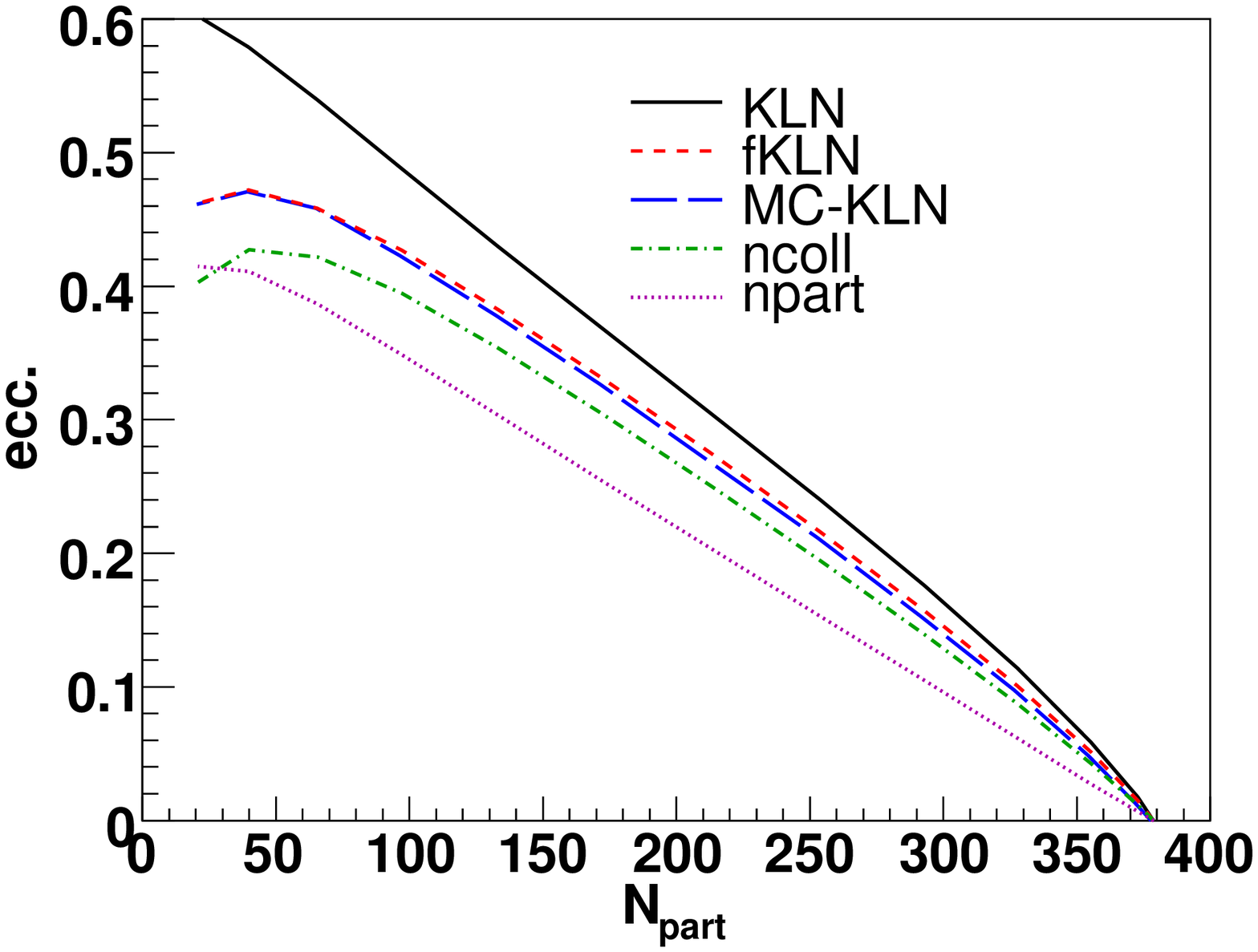}
\caption{(Color online) The eccentricity $\varepsilon$ in Au+Au
  collisions predicted by the various models. fKLN and MC-KLN give
  almost identical results. The MC-KLN result was obtained using the
  oversampling technique with $n=100$, which corresponds roughly to a
  delta function sampling profile, just as used in the fKLN model.  }
\label{fig:ecc_TA_MC}
\end{center}
\end{figure} 

First, we show the centrality dependence of the
multiplicity from the MC-KLN model in Fig.~\ref{fig:npart_TA_MC}
together with the result of the original KLN approach.  For
clarity, we have not adjusted the overall normalization to the
most central events. The multiplicity as a function of the centrality
from MC-KLN is very similar to the original KLN and to the fKLN approach.
The overall normalization is about 5\% lower. The reason is that the
multiplicity as a function of the density has negative curvature
(second derivative), so fluctuations reduce the multiplicity.

In Fig.~\ref{fig:ecc_TA_MC}, we compare the eccentricity among the
various approaches.  Since the fKLN result was obtained using a delta
function sampling profile (by using $T_A$ instead of
$\widetilde{T}_A$), we employed the oversampling technique with a
factor $n=100$ for the MC-KLN model. This way, the two models use the
same sampling profile and the results are comparable. It can be seen
that the eccentricity from MC-KLN is almost identical to the fKLN
scaling, which is slightly higher than $n_{\rm coll}$ scaling.  Note
that any additional fluctuation (e.g. in the evolution of the
BFKL-ladders) would not influence the standard eccentricity
(\ref{eq:ecc}), which is due to its definition; the denominator and
the numerator are averaged separately and both are linear functions of
the weight.

The eccentricity from the CYM approach~\cite{LV06} behaves
similar to the number of collision scaling in the
Glauber model.  Note that the CYM approach in Ref.~\cite{LV06} does not
include effects of nucleon fluctuations. Hence, we
conclude that the main reason for the reduced eccentricity as
compared to the original KLN approach is the use of the thickness
functions in the definition of the saturation scale as argued in
Ref.~\cite{LV06}. Both, the CYM approach from Ref.~\cite{LV06} and our
fKLN approach, however, predicted larger $\varepsilon$ than the
Glauber model for soft collisions where $dN/dyd^2{\bf r}_\perp$ scales
with the density of participants $n_{\rm part}$.

\subsection{Eccentricity Fluctuations}

We now turn to a comparison of various definitions of
the eccentricity which are relevant for the scaling properties of the
 experimental data for $v_2$. Since we now consider event-by-event
variables, we cannot use oversampling any more. This changes the
results somehow, since the interaction range of the nucleon-nucleon
interaction plays some role. In Appendix \ref{appendixA}, we summarize
how to include the interaction range in the numerical computations of
the number of participants and collisions.

First we define the participant eccentricity $\varepsilon_\mathrm{part}$
according to Ref.~\cite{Manly:2005zy}
in order to account for fluctuations in the directions
of the major axes of the overlap region
and of the ``center of gravity'',
\begin{equation}
\varepsilon_\mathrm{part}
= \frac{\sqrt{(\sigma^2_y - \sigma^2_x)^2+4\sigma^2_{xy}}}
       {\sigma^x_y+\sigma^2_x},
\end{equation}
where $\sigma_x$ and $\sigma_y$ are the RMS widths of the participant
nucleon distribution projected on the $r_x$ and $r_y$ axes, respectively,
and $\sigma_{xy}=\langle xy \rangle - \langle x\rangle\langle y\rangle$.
We also compute 
\begin{equation}
  \varepsilon\{2\} \equiv \sqrt{\langle\varepsilon^2_\mathrm{part}\rangle}
\end{equation}
which is closely related to the event-plane method of the measurements
of elliptic flow. For four-particle cumulants of azimuthal correlations
~\cite{Bhalerao:2006tp} we have:
\begin{equation}
  \varepsilon\{4\} \equiv \left(
    2\langle\varepsilon^2_\mathrm{part}\rangle^2
    -\langle\varepsilon^4_\mathrm{part}\rangle \right)^{1/4}.
\end{equation}

\begin{figure}[tb]
\begin{center}
\includegraphics[width=8cm,clip]{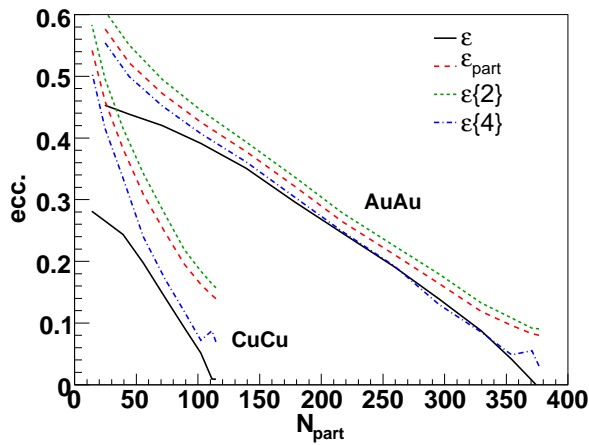}
\caption{(Color online) Various eccentricities for the MC-KLN model.
  No oversampling was used, which corresponds to a finite range
  sampling profile.}
\label{fig:ecc_e4}
\end{center}
\end{figure} 
In Fig.~\ref{fig:ecc_e4} we show $\varepsilon_\mathrm{part}$,
$\varepsilon\{2\}$, $\varepsilon\{4\}$ together with the standard
$\varepsilon$ from the MC-KLN model, for Au+Au and Cu+Cu collisions at
$\sqrt{s}=200$~ GeV, using again the energy density as weight in the
eccentricities. Compared to Fig.~\ref{fig:ecc_TA_MC}, the standard
eccentricity is reduced somewhat, since we did not use the oversampling
technique in this plot (see also Fig.~\ref{fig:ecc_glauber} of Appendix A).
\Red{As observed before
\cite{Manly:2005zy,Voloshin:2006gz,Ollitrault:1992bk}
in the Glauber model approach,
$\varepsilon\{2\}$ deviates most from the standard definition, and
the deviation is much larger for smaller systems. Since
the source of fluctuations in this work is the position of the nucleons,
the influence of the fluctuations on the results is very similar to that
of the Glauber model approach.
}

\begin{figure}[tb]
\begin{center}
\includegraphics[width=8cm,clip]{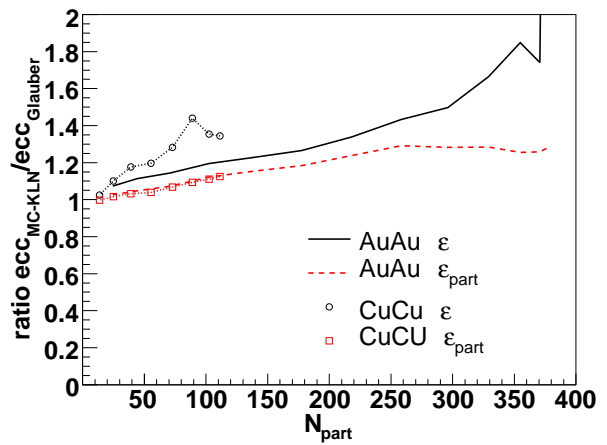}
\caption{(Color online) The standard eccentricity $\varepsilon$ and $\varepsilon_{\mathrm part}$
  of MC-KLN relative to the Glauber model result.  }
\label{fig:ecc_ratio}
\end{center}
\end{figure} 

\Red{A detailed comparison to the Glauber eccentricity can be seen in
Fig.~\ref{fig:ecc_ratio}, where we show the ratio of the eccentricity
obtained by the MC-KLN model to the eccentricity of the Glauber model.
In Au+Au collisions, we see an increase of the ratio
$\varepsilon_{\mathrm {MC-KLN}}/\varepsilon_{\mathrm{Glauber}}$ by
more than 50\% for central and 10\% for peripheral collisions.
Comparing to Cu+Cu collisions, we notice a discrepancy of the ratios
for a system size above 50 participants.  Using the
$\varepsilon_{\mathrm part}$ definition, which accounts for
fluctuations of the major axes, we observe that both systems follow
the same curve and the enhancement is not more than 30\%.  Therefore,
precise experimental data would be able to distinguish MC-KLN from the
usual Glauber scaling by examining the scaling properties of the
elliptic flow.}

So far we assumed that additional fluctuations (e.g. from the
evolution of the BFKL-ladders) in the gluon production are
negligible. While this is true for the standard definition of the
eccentricity, it does not necessarily hold for event-by-event
definitions such as $\varepsilon_\mathrm{part}, \varepsilon\{2\}$ and
$ \varepsilon\{4\}$. In these cases, the eccentricity might increase.
We tested this assumption by including simple Poissonian fluctuations
in the gluon number but found that $\varepsilon\{2\}$ increased by
only about $0.02$.

\section{Discussion and conclusions}

\begin{figure}[tb]
\begin{center}
\includegraphics[width=8cm,clip]{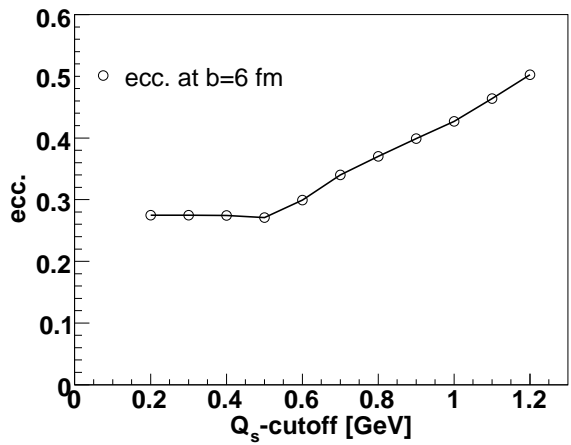}
\caption{
The eccentricity as a function of the $Q_{s,min}$ low-density
  cutoff.}
\label{fig:ecc_thr}
\end{center}
\end{figure} 

We developed a simple model which includes fluctuations of the
positions of the nucleons in the $k_\perp$-factorization 
formula. It naturally leads to a gluon production cross section which
respects factorization. The
multiplicity as a function of centrality does not change much as
compared to the original KLN approach but the
predicted eccentricity is somewhat lower.

\label{sec:ecc_thr}
It should be noted, however, that the gluon distribution calculated
here corresponds to very early times on the order of
$\sim1/Q_s$. Therefore, it is not necessarily identifiable with the
initial condition for hydrodynamics since it could be modified during
the thermalization stage. A detailed analysis requires simulations of
the time evolution from the early production time $1/Q_s$ until
thermalization. This could be done within parton cascade
models~\cite{molnar01,Xu:2004mz} which should be coupled to the
Yang-Mills equations describing the soft gluons~\cite{NaraCPIC}.
This is beyond the purpose of the present paper.

Our qualitative expectation is that the low-density surfaces of the
overlap zone may not thermalize (and should therefore not be included
in the initial condition for hydrodynamics), while the gluons produced
closer to the center do. We can assess the consequences of such a
simple picture by increasing the minimal saturation scale
$Q_{s,min}(x_i)$ which acts as a cutoff. This amounts to removing the
very high-$p_\perp$ gluon jets as well as the low-density part of the
bulk near the surface. The effect on the eccentricity is shown in
Fig.~\ref{fig:ecc_thr}.  One observes that the eccentricity increases
significantly with the cutoff.  The fact that $\varepsilon$ is flat up
to $Q_{s,min}=0.5$~GeV is due to the construction of the fKLN uGDF:
the saturation scale never drops below that for a single nucleon and
smaller cutoffs therefore do not matter. Since no thermalization is
expected to occur in $p+p$ or $p+A$ collisions, the relevant cutoff
for the hydrodynamic initial condition might be well above
$\sim0.5$~GeV. It would be interesting to study the hydrodynamical
evolution of various initial conditions in order to check how
sensitive the elliptic flow responds.

Another advantage of the fKLN uGDF is that it reproduces the correct
scaling of the multiplicity with $n_{\rm ncoll}=\sigma T_A T_B$ at
high $p_t$ in the DGLAP regime. This is not the case for the original
KLN ansatz. There, the high-$p_t$ multiplicity also scales with
$Q_{s,A}^2 Q_{s,B}^2$, but with $Q_{s,2} \sim \npart$, one does not
recover scaling with $T_A T_B$. As we have shown, the naive
modification to $Q_s^2\sim T_A$ does not help, since the saturation
momentum at the edge of a nucleus drops below that for a single
nucleon, and the $k_\perp$-factorization formula does not reproduce
the centrality dependence of the multiplicity.  Our fKLN (and MC-KLN)
uGDF solves these problems simultaneously: while it fits correctly the
measured centrality dependence of the multiplicity, it does not break
factorization, the nuclear saturation scale is bound from below by the
saturation scale of a single nucleon, and it reproduces the correct
$n_{\rm coll}$ scaling at high $p_t$.

\begin{appendix}

\section{Number of participants and Interaction profiles}
\begin{figure}[tb]
\begin{center}
\includegraphics[width=8cm,clip]{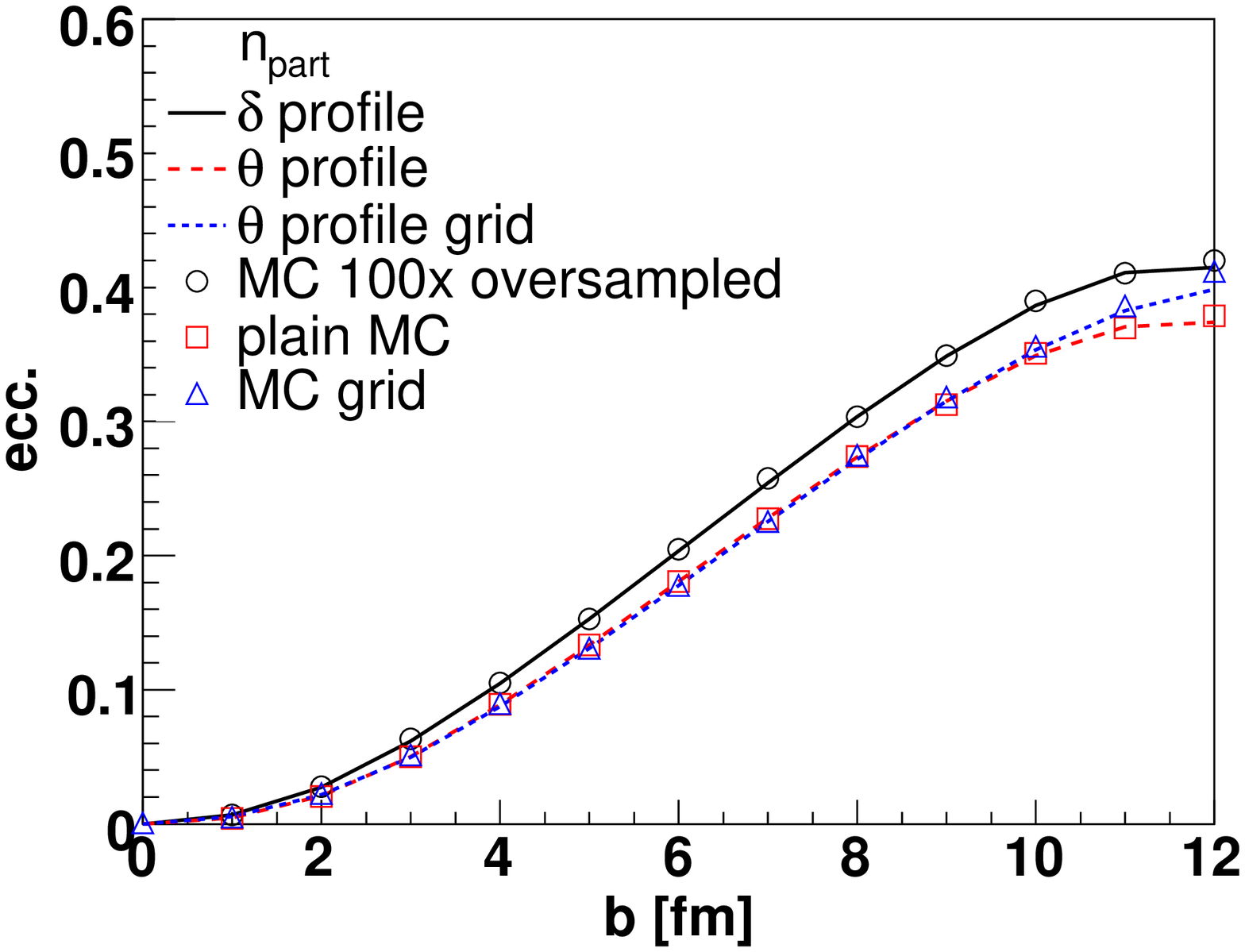}
\caption{(Color online) Eccentricity for the Glauber model with number of
  participants scaling of $dN/d^2{\bf r}_\perp$ obtained with the
  following methods: delta function interaction profile (point
  nucleons), step function (hard sphere nucleons) and step function
  with sampling on a grid.
}
\label{fig:ecc_glauber}
\includegraphics[width=8cm,clip]{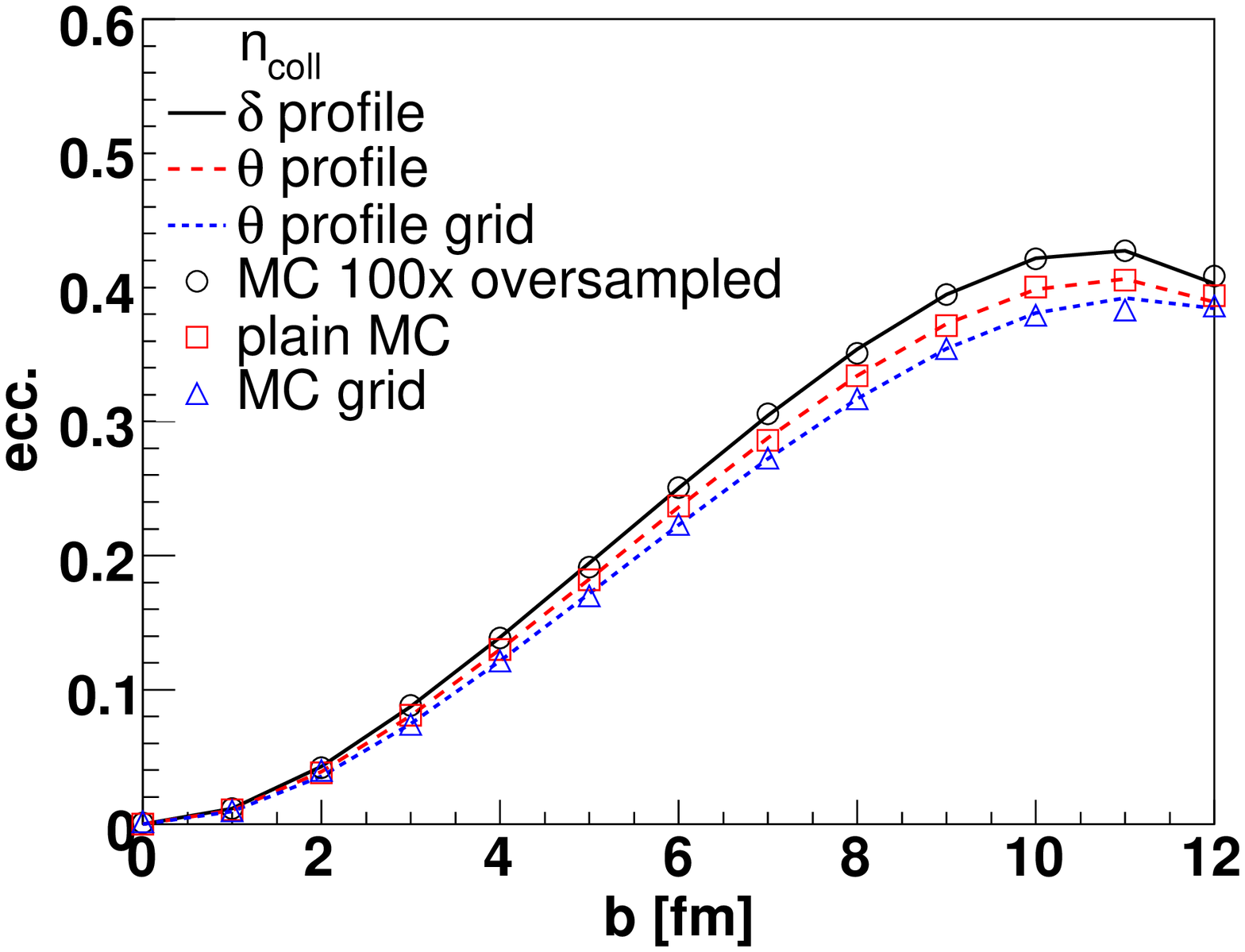}
\caption{(Color online)
Same as Fig.~\ref{fig:ecc_glauber} for number of collisions scaling.}
\label{fig:ecc_ncoll}
\end{center}
\end{figure} 

\label{appendixA}
\subsection{Delta function interaction profile}
The densities of participants in the transverse plane are given by
\begin{eqnarray}
n_{{\rm part},A} &=& T_A \left( 1-\left(1-\sigma {T}_B/B\right)^B\right)  \\
n_{{\rm part},B} &=& T_B \left( 1-\left(1-\sigma {T}_A/A\right)^A\right)  \\
n_{\rm coll} &=& \sigma T_A {T}_B ~.
\end{eqnarray}
These definitions assume a delta function interaction profile
$f(r)=\sigma \delta(r)$. In a Monte Carlo model, one can approximate
the delta function via the oversampling technique: $\sigma \rightarrow
\sigma/n$ and $A\rightarrow A n$, attributing a weight $1/n$ to each nucleon.

\subsection{Finite range interaction profile}

Next, we discuss how to calculate the number of participants and the
number of collisions respecting finite range interaction. The thickness
function is convoluted with the interaction profile, 

\begin{eqnarray}
\widetilde{T}_A({\bm r}_\perp) &=& \int d^2{\bm r}'_\perp \frac{ f({\bm r}_\perp-{\bm r}'_\perp)}{\sigma} T_A({\bm r}'_\perp) \nonumber ~.
\end{eqnarray}
Therefore, we have
\begin{eqnarray}
\label{eq:A4}
n_{{\rm part},A} &=& T_A \left( 1-(1-\sigma \widetilde{T}_B/B)^B\right)  \\
\label{eq:A5}
n_{{\rm part},B} &=& T_B \left( 1-(1-\sigma \widetilde{T}_A/A)^A\right)~.
\end{eqnarray}
The number of collisions can be derived in a similar way. Special
attention is needed when specifying the location of the collision:
$n_{\rm coll}= \sigma T_A \widetilde{T}_B$ locates the collision at
the participating nucleon from nucleus A and $n_{\rm coll}= \sigma T_B
\widetilde{T}_A$ locates the collision at the nucleon from nucleus
$B$. The following definition
\begin{eqnarray}
\label{eq:A6}
n_{\rm coll}=\frac{\sigma }{2}(T_A\widetilde{T}_B+\widetilde{T}_A T_B )
\end{eqnarray}
uses the center-of-mass position and is symmetric under the exchange
$A \leftrightarrow B$.

The definitions (\ref{eq:A4}-\ref{eq:A6}) agree with a straightforward
Glauber Monte Carlo, where the positions of the nucleons are sampled
according to the Woods-Saxon distribution, and where the profile functions
determine whether an inelastic interaction for a given pair of nucleons
occurs.

\subsection{Monte Carlo Method on a grid}

Sampling the number of participants within the interaction range
corresponds to using the convoluted thickness functions
for both nuclei. The corresponding definitions for the
participant and collision densities are:
\begin{eqnarray}
n_{{\rm part},A} &=& \widetilde{T}_A \left( 1-(1-\sigma \widetilde{T}_B/B)^B\right)  \\
n_{{\rm part},B} &=& \widetilde{T}_B \left( 1-(1-\sigma \widetilde{T}_A/A)^A\right) \\
n_{\rm coll}&=&{\sigma } \widetilde{T}_A\widetilde{T}_B 
\end{eqnarray}
Here the cross section is used in two ways: once for determining
the interaction, and once for sampling the nucleons over the
transverse area given by the cross section. Strictly speaking, this
implies double-counting but this method provides good approximations to
eqs.(\ref{eq:A4}-\ref{eq:A6}), as shown below.

Figs.~\ref{fig:ecc_glauber} and \ref{fig:ecc_ncoll} show the
eccentricity for number of participants and number of collisions
scaling obtained with different interaction profiles and sampling
methods. The normal definition of $T_A$ corresponds to a delta
function interaction profile. It is reproduced by a Monte Carlo
oversampling technique. A different interaction profile (in this case,
a $\theta$-function: $f(r)=\theta(\sqrt{\sigma/\pi}-r)$) leads to
formulas (\ref{eq:A4}-\ref{eq:A6}).  The corresponding Monte Carlo
method (denoted as plain MC) is straightforward: the profile function
determines the probability for an interaction for each
nucleon-pair. The third method, Monte Carlo on a grid, samples the
nucleon densities within the interaction range. It is very similar to
plain Monte Carlo. However, the number of collision scaling is
somewhat smaller.

\end{appendix}

\begin{acknowledgments}
We are indebted to Adrian Dumitru for illuminating
discussions and encouragement.  We thank M.~Gyulassy for useful
comments, and T.~Lappi and R.~Venugopalan for interesting discussions.
YN acknowledges support from DFG. HJD acknowledges support from
BMBF grant 05 CU5RI1/3.
\end{acknowledgments}

\end{document}